\documentclass[prd,aps,preprint,amsmath,nofootinbib,amssymb,eqsecnum,showkeys,tightenlines]{revtex4-1}
\usepackage{slashed}
\usepackage{epsfig,latexsym,cancel,amssymb,amsmath,verbatim,mathrsfs}
\usepackage{color}
\usepackage{graphicx}

\def\ra{\rightarrow}
\def\L{\left(}
\def\R{\right)}
\def\wt{\widetilde}
\def\Ld{\Lambda}
\def\ld{\lambda}
\def\f{\frac}
\newcommand{\be}{\begin{equation}}
\newcommand{\ee}{\end{equation}}
\newcommand{\bea}{\begin{eqnarray}}
\newcommand{\eea}{\end{eqnarray}}
\newcommand{\ba}{\begin{array}}
\newcommand{\ea}{\end{array}}

\long\def\symbolfootnote[#1]#2{\begingroup%
\def\thefootnote{\fnsymbol{footnote}}\footnote[#1]{#2}\endgroup}

\newcommand{\eq}[1]{Eq.~\eqref{#1}}

\newcommand{\beq}{\begin{equation}}
\newcommand{\eeq}{\end{equation}}

\begin{document}

\title{Minimal Dark Matter in the Local $B-L$ Extension}

\author{Chengfeng Cai}
\email[E-mail: ]{ccfspesysu@gmail.com}
\affiliation{School of Physics, Sun Yat-Sen University, Guangzhou 510275, China}

\author{Zhaofeng Kang}
\email[E-mail: ]{zhaofengkang@gmail.com}
\affiliation{School of physics, Huazhong University of Science and Technology, Wuhan 430074, China}

\author{Hong-Hao Zhang}
\email[E-mail: ]{zhh98@mail.sysu.edu.cn}
\affiliation{School of Physics, Sun Yat-Sen University, Guangzhou 510275, China}

\author{Yupan Zeng}
\email[E-mail: ]{zengyp8@mail2.sysu.edu.cn}
\affiliation{School of Physics, Sun Yat-Sen University, Guangzhou 510275, China}

\date{\today}

\begin{abstract}

The minimal gauge group extension to the standard model (SM) by the local $U(1)_{B-L}$ (MBLSM) is well known as the minimal model to understand neutrino mass origins via the seesaw mechanism, following the gauge principle. This ``small" symmetry also has deep implication to another big thing, dark matter (DM) stability. We demonstrate it in the framework of minimal dark matter (MDM), which aims at addressing two basic questions on DM, stability and the nature of interactions. However, stability and perturbativity may only allow the fermionic quintuplet. The situation is very different in the MBLSM, which leaves the subgroup of $U(1)_{B-L}$, the matter parity $(-1)^{3(B-L)}$, unbroken; it is able to stabilize all of the weakly-interacting {MDM candidates } after assigning a proper $U(1)_{B-L}$ charge. For the candidates with nonzero hypercharge, the phenomenological challenge comes from realizing the inelastic DM scenario thus evading the very strict DM direct detention bounds. We present two approaches that can slightly split the CP-even and -odd parts of the neutral components: 1) using the dimension 5 operators, which works for the $U(1)_{B-L}$ spontaneously breaking at very high scale; 2) mixing with {other fields} having zero hypercharge, which instead works for a low $U(1)_{B-L}$ breaking scale.

\end{abstract}


\maketitle

\section{Introduction}

From the theoretical side, the fundamental properties of the dark matter (DM) particle are of priori importance. For instance, why is DM stable? What kind of interactions that make DM acquire the correct relic density? and so on. They guide us to explore DM models that may shed light on these questions. The minimal dark matter (MDM) is an excellent example~\cite{Cirelli:2005uq,Cirelli:2007xd}, where DM is assumed to be the neutral component of the larger representation of $SU(2)_L$ assigned a proper $U(1)_Y$ charge. It answers those questions as the following:
\begin{itemize}
\item As long as the representation is sufficiently large, the MDM is supposed to be automatically long-lived without invoking any artificial DM protecting symmetry. Given the standard model (SM) field content along with the MDM field extension, the gauge and spacetime symmetries accidentally forbid the dangerous operators that lead to MDM decay. Such a situation is similar to {forbid the proton decay} in the SM.
\item It perfectly represents the most popular DM candidate, the weakly interacting massive particles (WIMP). The electroweak (EW) interactions should be the dominant interactions that the MDM participate in; they determine the relic density of MDM, hence DM mass to be multi-TeV.
\end{itemize}
The lowest dimensional MDM successfully answer both are a quintuplet and septuplet for the spin-1/2 and spin-0 fields, respectively.

We may have to pay special attentions on the spin-0 MDM candidates. The septuplet scalar cannot be the original MDM candidate because it admits dimension 5 operators {which were missed before,} could render the MDM short lived~\cite{MDM:d=5}. For the more general cases, the perturbativity requirement yields a strong constraint. Ref.~\cite{LP} pointed out that the self-couplings of scalar MDM, which again were missed before, run fast and hit the Landau pole at a scale far below the scale of grand unification theories (GUT), not mentioning to the Planck scale.~\footnote{Introducing fermions forming Yukawa coupling with the scalar MDM may relax this problem~\cite{Cai:2015kpa}.} On the other hand, a lot of people give up the accidental stability of MDM but just explore phenomenologies of the (purely) weakly-interacting WIMP DM with lower representations, at the price of imposing DM parity by hand~\cite{Hambye:2009pw,Arhrib:2013ela,Araki:2011hm,
AbdusSalam:2013eya,Cheung:2013dua,
Cai:2017wdu,Xiang:2017yfs}. We call them the generalized MDM, where ``minimal" refers to the number of parameters.

Therefore, it is the right time to modify the setup of the MDM, for the sake of maintaining its original merits, accidental stability, purely weakly-interacting and perturbative up to some putative fundamental scale. Viewing from the first item, also the most important one, extending the SM gauge group is a reasonable option. The minimal extension nevertheless having strong motivations is the local $U(1)_{B-L}$ (dubbed as MBLSM), which provides an elegant framework to understand the nonvanishing neutrino masses: Three right-handed neutrinos (RHNs) are introduced to fulfil anomaly cancellation and then the seesaw mechanism~\cite{seesaw} can be naturally realized with the seesaw scale tied to the $U(1)_{B-L}$ spontaneously breaking scale. Previously, the much larger group structure, namely the left-right symmetric groups $SU(2)_L\times SU(2)_R\times U(1)_{B-L}$ has been considered~\cite{Heeck:2015qra,Garcia-Cely:2015quu,
Berlin:2016eem,Cao:2017rjr}.
An alternative approach without gauge group extensions is assigning eccentric hypercharge to the MDM, resulting in millicharge MDM~\cite{DelNobile:2015bqo}.

In this paper we show why the MBLSM is an attractive setup for the MDM. {In a sense, due to the quantum gravity effects, only those discrete symmetries which originate from some continuous gauge symmetries can survive in the low energy effective theory~\cite{Ibanez:1991hv,Krauss:1988zc}.} As a matter of fact, the local $U(1)_{B-L}$ being helpful to stabilize both visible and invisible matters in the supersymmetric models has been known for a long time. The matter parity $P_M\equiv(-1)^{3(B-L)}$ forbids the dangerous baryon and lepton number violating operators at the same time~\cite{PM}, thus removing decay channels both for the proton and the lightest sparticle. $P_M$ can be a remanent of the local $U(1)_{B-L}$, provided that it is spontaneously broken by the vacuum expectation value (VEV) of a scalar field with an even integer value of $3(B-L)$~\cite{Ibanez:1991hv,Martin:1992mq,Martin:1996kn}. Essentially, the above idea {is irrelevant} to supersymmetry, so it is of great interest to extend it to the non-supersymmetric models. Several pioneered works along this line are grounded on the ambitious setup, namely the GUT~\cite{Kadastik:2009dj,Mambrini:2015vna,Frigerio:2009wf}. But our setup is less ambitious, the MBLSM. It incorporates three RHNs and as well as one $U(1)_{B-L}$ Higgs $\Phi$, whose $U(1)_{B-L}$ charge gives $3(B-L)=6$, an even integer hence leaving the subgroup $P_M$ unbroken.~\footnote{This is not the whole story of making use of the local $B-L$ to guarantee DM stability, e.g., scalar DM can be stabilized by an accidental $Z_3$ instead of $P_M$~\cite{Guo:2015lxa}. In addition, there are also studies based on the unconventional $U(1)_{B-L}$ charge assignment~\cite{Patra:2016ofq}.} Then, it opens a much wider MDM model space than the original one. Particularly, the candidates dwelling in the low dimensional representations now turn to be excellent samples. They show good detection prospects at all kinds of detectors, because their masses are not driven to a couple of TeVs. We concentrate on the scalar MDM with nonzero hypercharge, which arguably must scatter with nucleons inelastically to avoid the very strong exclusions from the DM direct detections. We present the mechanisms to generate the inelastic MDM~\cite{TuckerSmith:2001hy}, which have a close relation with the breaking scale of $U(1)_{B-L}$.

The paper is organized as the following: In Section \ref{sect2} we explain how the local $U(1)_{B-L}$ could stabilize MDM in the MBLSM and present ways to realize inelastic MDM, focusing on the spin-0 cases. In Section \ref{sect3} we briefly comment on the constraints from the current experiments. Section \ref{sect4} contains the conclusions and discussions.

\section{MDM in the minimal $SU(2)_L\times U(1)_Y\times U(1)_{B-L}$ model }\label{sect2}

We first introduce the convention. A field is labelled by its $SU(2)_L\times U(1)_Y\times U(1)_{B-L}$ quantum numbers $(2j+1, Y/2, B-L)$ with $j=0, 1/2, 1...$, and its $2j+1$ components are labelled by their QED charges $T_3+\f{Y}{2}$: $(j+\f{Y}{2}, j+\f{Y}{2}-1,...,-j+1+\f{Y}{2},-j+\f{Y}{2})$. The (generalized) MDM candidate is the neutral component; the term ``generalized" will be dropped hereafter for short. To facilitate constructing complete gauge invariant interactions, we write them in $SU(2)_L$ tensors, whose components correspond to the components listed above, for example, two representations which will be studied later:
\begin{eqnarray}\label{}
T_{(3,-1/2)}=\L\begin{array}{c}
  T^{0} \\
   T^{-} \\
 T^{--}
\end{array}\R
=\L\begin{array}{c}
  T^{11} \\
  \sqrt{2}T^{12} \\
  -T^{22}
\end{array}\R,
\quad
\Sigma_{(4,-1/2)}=\L\begin{array}{c}
  \Sigma^{+} \\
   \Sigma^{0} \\
 \Sigma^{-}\\
 \Sigma^{--}
\end{array}\R
=\L\begin{array}{c}
  \Sigma^{111} \\
  \sqrt{3}\Sigma^{112} \\
  \sqrt{3}\Sigma^{122}\\
  \Sigma^{222}
\end{array}\R,
\end{eqnarray}
where the $U(1)_{B-L}$ charge is irrelevant.

Stability, purity (almost purely weakly-interacting), perturbativity and safety restrict the MDM candidates in the MBLSM. With the help of $U(1)_{B-L}$, it is ready to ensure the stability of MDM candidates, even those having a low $j$. But now the presence of new gauge interaction may change the interactions of MDM. In order to make it still mainly be weakly interacting, we require that $U(1)_{B-L}$ is broken at a very high scale far above the TeV scale, otherwise its gauge coupling should be fairly small. {Either condition} can be easily satisfied, and we will see that the former condition is even necessary for the inelastic scalar MDM via high dimensional operators. The requirement that the model should not hit the Landau pole at least below the GUT scale is strong, it negates scalar MDM with $j\geq 2$~\cite{LP}. The last but not the least, the MDM has to escape the strong DM exclusion limits. The fully weakly-interacting DM scattering with nucleons having $Z$-boson contribution has definitely been ruled out by DM direct detection experiments such as PandaX-II~\cite{Cui:2017nnn}. Then MDM must be either self-conjugate~\footnote{Our exact meaning is that they do not carry hypercharge. However, in the presence of $U(1)_{B-L}$, ``self-conjugate" is not equivalent to ``zero hypercharge". For the historical reason, we still regard { them} equivalent throughout this paper. It is justified if the $U(1)_{B-L}$ gauge interactions are decoupled.} or quasi self-conjugate (namely inelastic) so as to eliminate/suppress the spin-independent DM-nucleon elastic scattering mediated by $Z$. In the following subsections we will expand them point by point.

\subsection{Stability: $V_p$-odd MDM}

\begin{table}[]
\begin{center}
\begin{tabular}{|c|c|c|c|c|}
\hline
~ ~~ ~~ ~~ ~&~~\rm spin~ ~& ~~$SU(2)_L$ ~~& ~~$U(1)_Y$~~ & ~~ $U(1)_{B-L}$~~  \\
\hline
$L$ & $1/2$ &     2  &    -1/2  &   -1 \\
\hline
$e_R$ &$1/2$ &     1&    1 &   -1   \\
\hline
$H$ &$0$ &   2  &    1/2 &   0   \\
\hline\hline
$N_R$ &1/2 &     1  &    0  &    -1   \\
\hline
$\Phi$ &0 &     1  &    0  &   $ 2$ \\
\hline
\end{tabular}
\end{center}
\caption{Field content and quantum numbers in the minimal $U(1)_{B-L}$ extension; quarks which carry $B-L$ charge $1/3$ are not included. \label{QN}}
\end{table}%
In this subsection we show how $U(1)_{B-L}$ helps to stabilize MDM in the MBLSM. We list its field content in Table.~\ref{QN}. To break $U(1)_{B-L}$ and simultaneously give Majorana mass to RHNs, the $U(1)_{B-L}$ Higgs $\Phi$ carrying $3(B-L)$ charge $6$ should be introduced. Then, asides from the SM part, the Lagrangian of MBLSM includes the following terms
\begin{eqnarray}\label{seesaw}
-{\cal L}_{\rm MBLSM}\supset \overline{L} \wt HN_R+ \f{\ld_N}{2}\Phi \overline{ N_R^C} N_R+ V(\Phi),
\end{eqnarray}
where the $B-L$ Higgs sector $V_\Phi$ is responsible for breaking $U(1)_{B-L}$ at a scale $\langle \Phi\rangle \equiv v_\phi/\sqrt{2}$. As a consequence of the RHN Majorana mass terms from the condensation of $\Phi$, the $B-L$ charge of $\Phi$ is determined to be 2, which then leads to the remanent  $P_M=(-1)^{3(B-L)}$ after $U(1)_{B-L}$ spontaneously breaking. So, $P_M$, which has gauge origin, can be safely used to protect DM from decay. We would like to comment that a similar structure arises in the setup with $SU(2)_L\times SU(2)_R\times U(1)_{B-L}$ gauge groups~\cite{Heeck:2015qra,Garcia-Cely:2015quu,Berlin:2016eem}, where the $SU(2)_R$ triplet Higgs field accounting for the breaking $SU(2)_R\times U(1)_{B-L}\ra U(1)_{Y}$ also has 2 units of $U(1)_{B-L}$ charge, which again is due to the Majorana mass of RHNs.

Maybe it is more { ransparent} to consider the equivalent form of matter parity, namely the parity defined as $V_p=(-1)^{3(B-L)+2s}=(-1)^{2s}P_M$ with $s$ the spin of the particle; Lorentz invariance forces any interactions to give a factor $\Pi_i (-1)^{2s_i}=1$ and therefore a $P_M$ invariant theory is also $V_p$ invariant. In supersymmetry, this parity is the well known $R$-parity. All the SM particles are {$V_p$-even}.

The MDM candidates can be classified into two types, with or without $B-L$ charge. Let us first consider the latter case, where the MDM must be $P_M$ even, so seemingly it is not protected by $P_M$. This is true for a scalar MDM, for which $P_M$ is completely useless and consequently MDM stability again has to fall back on the high $j$ as in the original MDM setup, moreover, it may decay through a dimension-5 operator. Nevertheless, if the MDM is a fermion field, it would be $V_p$-odd; hence, it is automatically stable by virtue of $V_p$.~\footnote{Maybe it follow the spirit of MDM more closely than the $B-L$ charged MDM, for which one has to arrange the $B-L$ charge properly.} For instance, in the absence of $U(1)_{B-L}$, the fermionic MDM candidates with $j\leq 3/2$ fail because they admit decay modes by coupling to the operators listed in Table~\ref{decay}. However, all of these fatal operators are forbidden by $V_p$ or $U(1)_{B-L}$. As a result, in the MBLSM they have the potential to be good MDM candidates. But among them only $(3,0,0)$, which is weakly-interacting and self-conjugate, is the desired one. Whereas the others (in the form of Dirac pair to simply cancel anomalies), should mix with $(1,1,0)$ or $(3,0,0)$ to become (quasi) self-conjugate; see the example of $(3,1,0)$ and $(3,-1,0)$ Dirac pair in Ref.~\cite{Kang:2013wm}.

Now we turn our attention to the MDM candidates charged under $U(1)_{B-L}$. One unsatisfactory thing is that the charge assignments are arbitrary without further theoretical inputs, although this could trivially make the MDM stable, e.g., assigning a strange fractional charge, which refers to those not in the unit of $1/3$, to the MDM. Note that that kind of charge leads to ill definition of $V_p$. We will see that for the scalar MDM, the requirement of safety will select out the charge for those in need of mass splitting to be inelastic DM. The fermions neutral under $U(1)_{B-L}$ are $V_p$-odd, so we do not have particular motivations to consider the $U(1)_{B-L}$ charged fermionic MDM candidates. If appearing, they should have an even value of $3(B-L)$. There is an example in the supersymmetric $U(1)_{B-L}$ models~\cite{bilepton}, which is the superpartner of $\Phi$. But it is a SM singlet hence not the MDM candidate. Therefore, in what follows we will concentrate on the spin-0 MDM case. Actually, the discussions in two cases do not differ much, and comments on the fermionic MDM will be made if necessary.
 \begin{table}[]
\begin{center}
\begin{tabular}{|c|c|c|c|c|c|c|}
\hline
~ ~~ ~~ ~~ ~&~~$(1,0,0)$~ ~& ~~$\L 2,\f{1}{2},0\R$ ~~& ~~$(3,0,0)$~~ & ~~ $\L 3,1,0\R$~~& ~~ $\L 4,\f{1}{2},0\R$~~ & ~~ $\L 4,\f{3}{2},0\R$~~ \\
\hline\hline
\rm decay channel& $LH$ &     $He_R$  &    $LH$ &  $LH^*$  & $LHH^*$   &  $LHH $  \\
\hline
\end{tabular}
\end{center}
\caption{Fermionic MDM neutral under $B-L$; they have $j\leq 3/2$ and can decay into the fields listed in the second line if $U(1)_{B-L}$ does not act. For the three candidates with $j\leq 1$, they appear as the bino (singlino), Higgsino and wino, triplino~\cite{Kang:2013wm} in the supersymmetric models, respectively. \label{decay}}
\end{table}%

\subsection{ Safety: (quasi) self-conjugate MDM}\label{EOMS}

In this subsection we present two approaches to make the scalar MDM be (quasi) self-conjugate, namely to realize mass splitting between the CP-even and -odd parts of the neutral component in the field $(2j+1, Y/2, B-L)$, EOMS for short.

\subsubsection{$Y=0$ scalar MDM}

As argued before, the spin-0 MDM is putative to carry $U(1)_{B-L}$ charge odd times of 1/3. For clearness, we first pick out the special case with $Y=0$ thus already self-conjugate. They are $(1 ~{\rm or }~ 3,0,\pm \f{2n+1}{3})$ with $n=0,1,2...$, where $j\leq 3/2$ from perturbativity has been imposed. Decoupling the $B-L$ gauge interactions they are basically reduced to the complex singlet or triplet scalar DM with $Z_2$ parity, which is identified with $V_p$ here.

Among them, the one with $n=1$ will be useful to generate EOMS for the scalar MDM with $Y\neq 0$, so let us study their mass spectra in details for preparation. They are the singlet $\wt N: (1,0,-1)$, corresponding to the right-handed sneutrino DM~\cite{iIDM,Guo:2013sna,Cao:2017cjf} in the seesaw extended supersymmetric SMs, and the triplet $T: (3,0,-1)$. We focus on the non-trivial one, $T$, for illustration. Because $T$ is charged under $U(1)_{B-L}$, now it is a complex instead of real triplet MDM as in the usual case~\cite{Araki:2011hm}.~\footnote{To our knowledge, this example has not been discussed by other groups yet, except for the Ref.~\cite{Cai:2017wdu} which studied a phenomenological case mixing with a real singlet.} As a result, the additional term $\beta H_i^\dagger H^j  T^{ik}T^\dagger_{jk}$ is present. It, along { with} other terms especially the nontrivial term $A_T\Phi T^2$, generates the triplet mass spectra
  \begin{eqnarray}\label{}
      m_{T_R}^2&=&m_{T}^2+\f{\beta}{4}v^2+\sqrt{2}A_Tv_\phi,\cr
m_{T_I}^2&=&m_{T_R}^2-2\sqrt{2}A_Tv_\phi,
\cr m_{T_+}^2&=&m_{T_R}^2-\sqrt{2}A_Tv_\phi-
\f{1}{4}\sqrt{\beta^2v^4+32A_T^2 v_\phi^2},\cr
m_{T'_+}^2&=&m_{T_R}^2-\sqrt{2}A_Tv_\phi+
\f{1}{4}\sqrt{\beta^2v^4+32A_T^2 v_\phi^2},
\end{eqnarray}
which satisfy the tree-level mass sum rule: $m_{T_R}^2+m_{T_I}^2=m_{T_+}^2+m_{T'_+}^2$. The mass splitting between $T_R$ and $T_I$ is via the trilinear term { with dimensionful coupling}, so it can be significant even if $v_\phi$ is not large.

The charged component $T_+$ does not lie at the bottom of the spectra only in the limit $\beta=0$, an unwanted feature of the spectra. But radiative corrections may change this situation. Actually, it is well known that EW radiative corrections could lift the component with electronic charge $Q$ by an amount $0.17Q(Q+2Y/\cos\theta_w)$ GeV, compared to the one with $Q=0$~\cite{Cirelli:2005uq}. Note that this lift, in contrast to the mass shift due to the $\beta$-term which is suppressed by heavy $m_T$, is independent on $m_T$.~\footnote{This is true only at the one-loop level; at higher loop level the mass difference may develop a weak dependence on $m_T$~\cite{McKay:2017xlc}, which however is not important in our study.} Consider the hierarchy $\beta^2v^4\ll 32A_T^2 v_\phi^2\ll m_{T_R}^4$, one can make the following approximation
\begin{eqnarray}\label{order:T}
  m_{T_+}\approx m_{T_R}-\sqrt{2}\f{A_Tv_\phi}{m_{T_R}}-
  \f{1}{64\sqrt{2}}\f{\beta^2v^4}{A_Tv_\phi m_{T_R}},
  \end{eqnarray}
where $A_Tv_\phi>0$ is assumed and the opposite case leads to a similar result. So, $T_+$ is lighter than the lighter neutral component, $T_I$, by an amount measured by the last term of the right side of Eq.~(\ref{order:T}). If the radiative lift could offset it, one requires
 \begin{eqnarray}\label{}
  \f{1}{64\sqrt{2}}\f{\beta^2v^4}{|A_Tv_\phi|m_{T_R}}\lesssim 0.17
  {\rm GeV}\Rightarrow |\beta|\lesssim 0.46\L\f{|A_T v_\phi|}{10^4\rm GeV^2}\R^{1/2} \L\f{m_{T_R}}{\rm 5 TeV}\R^{1/2}.
  \end{eqnarray}
a fairly loosing bound. If instead $m_{T_R}^4\gg\beta^2v^4\ \gg 32A_T^2 v_\phi^2$, the second term on the right side of Eq.~(\ref{order:T}) vanishes and a somewhat more strict bound is obtained:
\begin{eqnarray}\label{}
|\beta|\lesssim 0.06\times \L m_{T_R}/5 \rm TeV\R.
  \end{eqnarray}
But anyway the $\beta$-term is not really troublesome here, and it is also true in the quadruplet discussed later on.

We would like to remind the readers that, even if EOMS is small, $Z$-boson does not mediate inelastic scattering between the triplet MDM and nucleons. The reason is traced back to its neutrality under $U(1)_Y$, and consequently there is no coupling like $T_0^\dagger \partial_\mu T_0 W_0^\mu$. The extra massive gauge boson $Z_{B-L}$ could do that job, but we are assuming that it is irrelevant to the MDM interactions.

\subsubsection{Splitting the $Y\neq0$ scalar MDM without mixing}

We first consider the single scalar MDM candidates with $Y\neq 0$. If their $U(1)_{B-L}$ charges are $-1$, they can be made quasi self-conjugate by means of coupling to the Higgs fields through higher dimensional operators. For $j\leq 3/2$, three candidates are allowed, and the corresponding higher dimensional operators are given by
\begin{eqnarray}\label{d=5}
 \f{1}{\Ld}\L\wt L H\R^2 \Phi,~~~~  \f{1}{\Ld^3}\L H T_{(3,-1,-1)} H\R^2 \Phi ,~~~~   \f{1}{\Ld}\L \Sigma_{(4,-1/2,-1)} H\R^2 \Phi.
\end{eqnarray}
One can explicitly check that a generic renormalizable potential $V(H,\Phi,X)$, with $X$ denoting the three scalar MDM fields, respects an accidental global $U(1)$ symmetry. It protects the mass degeneracy between the CP-even and odd parts of the neutral component of $X$. While the splitting operators in \eq{d=5} slightly break the accidental symmetry, generating EOMS thus the quasi self-conjugate MDM.

We take the cut-off scale $\Ld\sim M_{\rm Pl}$, while the $U(1)_{B-L}$ breaking scale $v_\phi$ is regarded as a free parameter, varying from the weak scale to the GUT scale.~\footnote{A very high scale $v_\phi$ always incurs the small coupling issue, namely requiring $\ld_{\phi X}\ll1$ in $\ld_{\phi X}|\Phi|^2|X|^2$ to allow a light $X$ at low energy. Supersymmetric models may be not subject to this issue.} If $v_\phi$ is very low, the effects from the higher order operators are completely negligible. For the splitting operators at dimension 5, the induced EOMS is estimated to be
\begin{eqnarray}\label{}
\delta m\sim \f{v_\phi}{\Ld} \f{v^2 }{M_{\rm DM}}\sim 10\times \L\f{v_\phi}{10^{13}\rm GeV}\R\L\f{10^{19}\rm GeV}{\Ld}\R \L\f{1\rm TeV}{M_{\rm DM}}\R \rm keV,
\end{eqnarray}
which fits the typical inelastic DM scenario. For the TeV scale MDM, $\delta m$ is at most a few 10 MeV, with the upper bound saturated for $v_\phi=M_{\rm GUT}$. That high scale $v_\phi$ does not incur a problem to the seesaw mechanism, as long as we require a relatively small coupling $\ld_N\lesssim 10^{-2}$ in \eq{seesaw}. As a generic feature of the EOMS via a dimension 5 operator, the $U(1)_{B-L}$ gauge interactions are completely irrelevant to the low energy phenomenologies. So, in this case decoupling $U(1)_{B-L}$ gauge interactions is justified in discussing the interactions of MDM. For EOMS above the keV scale, the heavier state in $X$ can three-body decay into a pair of neutrinos plus DM, mediated by a virtue $Z$ boson. The inverse decay width is very short {comparing to} the cosmic time scale. Hence it cannot be a relic today. As for the hypercharged triplet $T_{(3,-1,-1)}$, the corresponding splitting operator is at dimension 7. {Comparing} to the dimension 5 case, the resulted $\delta m$ suffers from one more suppression factor $\sim (v/\Ld)^2\sim 10^{-34}$. As a consequence, $T_{(3,-1,-1)}$ still fails to be a successful MDM candidate even if $v_\phi$ is close to the GUT scale. Therefore, only two candidates survive.

One is the inert doublet $\wt L$, which assembles the slepton in supersymmetry thus the notation here. This example has been extensively studied in the literatures by invoking an artificial $Z_2$ dark parity, well known as the inert Higgs doublet model~\cite{Arhrib:2013ela}. Our contributions, asides from giving the gauge origin of such a symmetry, is the prediction that the inert doublet DM must be an inelastic DM, with the exciting component having a mass at most a few MeV heavier than the ground state. Such a compressed inert Higgs doublet DM scenario has been studied in Ref.~\cite{Blinov:2015qva}, focusing on collider searches.

The other candidate is the quadruplet. We calculate the complete mass spectra in terms of the generic Higgs potential $V(\Sigma, H, \Phi)$, dropping the subscripts of the MDM fields to avoid clumsy. Besides the dimension 5 operator given in \eq{d=5}, like the triplet case $V(\Sigma, H, \Phi)$ contains a $\beta$-term, $\beta H_i^\dagger H^j  \Sigma^{ikl}\Sigma^\dagger_{jkl}$, which is able to split the components with different charges. Then the resulting mass spectra is given by
\begin{eqnarray}\label{}
      m_{\Sigma_R}^2&=&m_{\Sigma}^2-\f{\sqrt{2}\gamma}{3}v^2,\cr
m_{\Sigma_I}^2&=&m_{\Sigma_R}^2+\f{2\sqrt{2}\gamma}{3}v^2,
\cr m_{\Sigma_+}^2&=&m_{\Sigma_R}^2+\f{\sqrt{2}\gamma}{3}v^2+
\f{v^2}{6}\sqrt{\beta^2+6 \gamma^2},\cr
 m_{\Sigma'_+}^2&=&m_{\Sigma_R}^2+\f{\sqrt{2}\gamma}{3}v^2-
\f{v^2}{6}\sqrt{\beta^2+6 \gamma^2},
\cr m_{\Sigma_{++}}^2&=&m_{\Sigma_R}^2+\f{\beta}{3}v^2,
\end{eqnarray}
we have defined $\gamma\equiv v_\phi/\Ld$, in order to compare with the results given in Ref.~\cite{AbdusSalam:2013eya}; see also Ref.~\cite{Cai:2017wdu}. Here $\gamma\ll1$, and thus $\beta\gg \gamma$ is supposed to hold. In this limit, one may readily check that either $\Sigma_{++}$ (for $\beta<0$) or $\Sigma'_+$ (for $\beta>0$) is the lightest component, rather than the neutral component. But again the radiative corrections could beat the $\beta$-term as long as $\beta$ is not large:
 \begin{eqnarray}\label{}
|\beta|\lesssim 0.3 \L \frac{Q}{2}\R^2\L \frac{M}{5{\rm TeV}}\R,
\end{eqnarray}
which is a loosing condition.

\subsubsection{Splitting the $Y\neq 0$ scalar MDM via mixing}

The second approach to make the scalar MDM with $Y\neq 0$ quasi self-conjugate is by mixing with those having $Y= 0$, namely the singlet $\wt N: (1,0,-1)$ and triplet $T: (3,0,-1)$; this approach accommodates EOMS at the renormalizable level. In this scenario, $v_\phi$ can be set to a fairly low scale as long as it is experimentally allowed. So, it does not have the problem stated in footnote 6, arising in the high dimensional operator approach. The minimal models involve two fields, concretely, the mixing singlet-doublet and singlet-triplet model (not listing the trivial and irrelevant terms),
\begin{eqnarray}\label{}
{\cal L}_{SD}&\supset&\L A_N\Phi \wt N^2+ A_{12}\wt L H \wt N^*+c.c.\R+\beta  H^i  \wt L^\dagger_{i} H_j^\dagger\wt L^j,\cr
{\cal L}_{ST}&\supset &\L A_N\Phi \wt N^2+ \ld_{13} HT^\dagger_{(3,-1,-1)}H\wt N+c.c.\R+\beta H_i^\dagger H^j  T_{(3,-1,-1)}^{ik}\L T_{(3,-1,-1)}\R^\dagger_{jk},
\end{eqnarray}
where the $\beta$-terms, unlike in the $(3,0,-1)$ case, merely contribute to the masses of the charged components; this feature may be useful to avoid the indirect detection bound (on $T_{(3,-1,-1)}$) discussed later. It is seen that in this approach the $B-L$ charge assignment is also fixed, to be one unit. We have recovered the subscript for the triplet, whose EOMS cannot be achieved via the higher dimensional operators. We will concentrate on these two typical examples, but the approach applies to other MDM candidates, for instance, the triplet-quadruplet model
\begin{eqnarray}\label{}
{\cal L}_{TQ}\supset {A_N}\Phi T_{(3,0,-1)}^2+ A_\Sigma T_{(3,0,-1)} \Sigma_{(4, 1/2,-1)}^* H+ c.c.
\end{eqnarray}
In addition, the above approach, through prolonging the mixing chain, can be generalized to the case with an even higher $j$ or larger hypercharge. As an example of existence, one can cascadingly generate EOMS for the quadruplet $\Sigma_{(4, -3/2, -1)}$, which fails via neither the higher dimensional operators nor the two-field chain owing to its higher hypercharge, now via the singlet-triplet-quadruplet (three-field) chain:
\begin{eqnarray}\label{}
{\cal L}_{STQ}\supset  A_N\Phi \wt N^2+ \ld_{13} HT^\dagger_{(3,-1,-1)}H\wt N+ A_\Sigma  \Sigma_{(4, -3/2,-1)} T_{(3,-1,-1)}^\dagger H+ c.c.
\end{eqnarray}
In principle, in this way all of the weakly-interacting particles can be made viable MDM candidates.

We first study the singlet-triplet mixing example. To make the DM not be dominated by the singlet $\wt N$, its mass $m_{\wt N}$ is supposed to be considerably heavier than the mixing partner. In this limit, the masses of the neutral components of the triplet are approximated to be
\begin{eqnarray}\label{}
m_{T_1}\approx m_T-\f{\ld_{13}^2}{32}\f{v^4}{\L m_{\wt N}^2+\sqrt{2}A_Nv_\phi-m_T^2\R m_T},\cr
m_{T_2}\approx m_T-\f{\ld_{13}^2}{32}\f{v^4}{\L m_{\wt N}^2-\sqrt{2}A_Nv_\phi-m_T^2\R m_T},
\end{eqnarray}
with the charged components having degenerate tree level mass $m_T$. While the two singlets have mass squared $m_{\wt N}^2\pm A_N v_\phi/\sqrt{2}$. The EOMS effect in the singlet sector is mediated to the triplet sector through the $\ld_{13}$-term, which is suppressed by the factor $v^2/m_{\wt N}^2\ll 1$, and consequently the EOMS in the triplet sector is insignificant:
\begin{eqnarray}\label{}
\delta m_{T}\approx \f{\ld_{13}^2}{16}\L\f{v}{m_{\wt N}}\R^4\f{\sqrt{2}A_Nv_\phi}{m_T}=0.2\L\f{\ld_{13}}{0.5}\R^2 \L\f{5\rm TeV}{m_{\wt N}}\R^4\L \f{2\rm TeV}{m_T}\R \L\f{A_N v_\phi}{4\rm TeV^2}\R\rm MeV,
\end{eqnarray}
assuming $\sqrt{2}A_Nv_\phi\ll m_{\wt N}^2$.

The singlet-doublet mixing model, where the massive coupling $A_{12}$ mediates the singlet EOMS, is capable of generating a mildly larger EOMS. Besides, the correct DM relic density allows the mass scale in the singlet-doublet model to be much lower than in the singlet-triplet model, which helps to further enhance EOMS. Similarly, one may get the masses of the two neutral components in $\wt L$:
\begin{eqnarray}\label{}
m_{\wt L_{1,2}}\approx m_{\wt L}-
\f{1}{16}\f{A_{12}^2v^2}{\L m_{\wt N}^2\pm A_Nv_\phi/\sqrt{2}- m_{\wt L}^2\R m_{\wt L}}.
\end{eqnarray}
{Again} in the heavy $m_{\wt N}$ limit, the resulting EOMS $\delta m_{\wt L}\simeq \f{1}{8}A_{12}^2v^2 A_Nv_\phi/(m_{\wt N}^4 m_{\wt L})$ can be around the GeV scale. This model was also studied in the left-right symmetric models~\cite{Garcia-Cely:2015quu}, where the $SU(2)_L$ singlet $\wt N$ is interpreted as a $SU(2)_R$ doublet, the partner of the $SU(2)_L$ doublet $\wt L$ for the sake of preserving the left-right parity.


\section{Possible constraints}\label{sect3}

The main purpose of this paper is making the generalized MDM maintain the main motivations of the original MDM, accidentally stable and purely weakly-interacting, but we can barely make new contributions to the MDM phenomenologies. They have been studied detailedly in Refs.~\cite{Cirelli:2007xd,Hambye:2009pw,Cirelli:2015bda,
Chun:2015mka}, from DM relic density to DM direct and indirect detections. In particular, the DM annihilations ${\rm DM+DM}\ra W^+W^-/ZZ$ and ${\rm DM+DM}\ra \gamma\gamma/Z\gamma$ in the galaxy, due to the Sommerfeld effect manifest at low velocity $v\sim 10^{-3}$, their cross sections could be greatly enhanced today, compared to those around the freeze-out epoch~\cite{Cirelli:2007xd,Chun:2015mka,Cirelli:2015bda,
Mitridate:2017izz}. Consequently, the TeV scale MDM candidates are stringently constrained by the MAGIC and FERMI-LAT continuum photon searches~\cite{FERMILAT}, which are sensitive to $\sigma (W^+W^-/ZZ)v_{rel}\sim {\cal O}(10^{-24})\rm cm^3/s$, and by H.E.S.S. gamma line searches~\cite{HESS}, which are sensitive to $\sigma (\gamma\gamma/Z\gamma)v_{rel}\sim {\cal O}(10^{-27})\rm cm^3/s$. They tend to rule out the $j\geq 1$ candidates, if they have very degenerate masses among all components, which is typical in our approaches given in subsection~\ref{EOMS}. But in the approach arriving EOMS via mixing, the mass splitting between the neutral and charged components might be significant, which helps to relax the constraints. Anyway, a more detailed analysis, including gamma ray/line constraints on the specific MDM candidates, should be done in a future publication.

The relatively new contribution may come from DM direct detections. The remarkable feature of our MDM candidates is that, for those MDM with $Y\neq0$, they tend to spin-independently scatter with nucleons inelastically. The scattering is mediated by the $Z$-boson, resulting in the cross section (for scalar DM-neutron scattering)
\begin{eqnarray}\label{}
\sigma_n=4\f{G_f^2m_n^2}{2\pi}Y^2=3.3\times 10^{-38}\L\f{Y}{1}\R^2\rm cm^{2}.
\end{eqnarray}
For such a large cross section, EOMS with $\delta m\lesssim 300$keV has been ruled out in the light of recent analysis using data of PICO~\cite{Bramante:2016rdh} and PandaX-II~\cite{Chen:2017cqc}. This already leaves the bulk space of the singlet-triplet model at the edge of exclusion.




\section{Conclusions and discussions}\label{sect4}

The MDM is supposed to give nice answers to two basic questions of DM, stability and interactions. However, the quintuplet fermion may be the only viable candidate confronting stability and perturbativity. On the other hand, the weakly-interacting particles are good cases in point of WIMP DM and they are widely studied in many contexts. We dub them the generalized MDM, which cannot address the stability mechanism and necessitates a DM protecting symmetry.

The MBLSM, extending the SM gauge groups minimally by the local $U(1)_{B-L}$, is a very attractive setup for the (generalized) MDM. Originally the MBLSM, following the gauge principle, is proposed to understand neutrino mass origin via the seesaw mechanism. It is found that the realization of the seesaw mechanism leaves the subgroup of $U(1)_{B-L}$, namely the matter parity $P_M=(-1)^{3(B-L)}$ (or equivalently $V_p=P_M (-1)^{2s}$) unbroken. Therefore, this gauged discrete symmetry potentially is powerful to stabilize the (generalized) MDM candidates. We find that it is able to stabilize all of the weakly-interacting particles given a proper $U(1)_{B-L}$ charge. For the candidates with $Y=0$, the phenomenological challenge is from realizing the inelastic DM scenario in order to evade the very strict DM direct detention bounds. We present two approaches to split the CP-even and -odd parts of the neutral components by:
\begin{itemize}
\item the dimension 5 operators, which works for the $U(1)_{B-L}$ spontaneously breaking at very high scale. More concretely, this approach applies only to $(2,-1/2,-1)$ and $(4,-1/2,-1)$.
\item  mixing with those having zero hypercharge, which works for a low $U(1)_{B-L}$ breaking scale. This approach applies to all of the {MDM} candidates, given a sufficiently long mixing chain.
\end{itemize}
Indirect detections from gamma rays impose a stringent bound on them, owing to the Sommerfeld effect. We need further detailed study to see if MDM with $j\geq 1$ still survive.

For the MDM candidates with $Y=0$ or realizing EOMS using the mixing MDM approach, the $U(1)_{B-L}$ gauge interactions can survive at low energy and may play a role. If they dominate, probably for the SM singlets which nevertheless are not MDM candidates, one may have similar phenomenologies studied in Ref.~\cite{Rodejohann:2015lca}.


\noindent {\bf{Acknowledgements}}
This work is supported in part by the National Science Foundation of China (11775086, 11375277).






\vspace{-.3cm}
\begin{thebibliography}{99}




\bibitem{Cirelli:2005uq}
  M.~Cirelli, N.~Fornengo and A.~Strumia,
    Nucl.\ Phys.\ B {\bf 753}, 178 (2006).


\bibitem{Cirelli:2007xd}
  M.~Cirelli, A.~Strumia and M.~Tamburini,
  Nucl.\ Phys.\ B {\bf 787}, 152 (2007).


\bibitem{MDM:d=5}
L. Di Luzio, R. Grber, J. F. Kamenik and M. Nardecchia, 
JHEP 1507, 074 (2015). 

\bibitem{LP}
Y. Hamada, K. Kawana and K. Tsumura,
Phys. Lett. B 747, 238 (2015).


\bibitem{Hambye:2009pw}
  T.~Hambye, F.-S.~Ling, L.~Lopez Honorez and J.~Rocher,
  JHEP {\bf 0907}, 090 (2009)
  Erratum: [JHEP {\bf 1005}, 066 (2010)]; 
  H.~E.~Logan and T.~Pilkington,
  Phys.\ Rev.\ D {\bf 96}, no. 1, 015030 (2017); 
  C.~El Aisati, C.~Garcia-Cely, T.~Hambye and L.~Vanderheyden,
  JCAP {\bf 1710}, no. 10, 021 (2017).

\bibitem{AbdusSalam:2013eya}
  S.~S.~AbdusSalam and T.~A.~Chowdhury,
  JCAP {\bf 1405}, 026 (2014).


\bibitem{Araki:2011hm}
  T.~Araki, C.~Q.~Geng and K.~I.~Nagao,
  Phys.\ Rev.\ D {\bf 83}, 075014 (2011); 
  Y.~Kajiyama, H.~Okada and K.~Yagyu,
  Nucl.\ Phys.\ B {\bf 874}, 198 (2013);  
  S.~S.~C.~Law and K.~L.~McDonald,
  JHEP {\bf 1309}, 092 (2013);
W.~B.~Lu and P.~H.~Gu,
  Nucl.\ Phys.\ B {\bf 924}, 279 (2017).


\bibitem{Arhrib:2013ela}
 L. Lopez Honorez, E. Nezri, J. F. Oliver and M. H. G. Tytgat, JCAP 0702, 028 (2007);
  A.~Arhrib, Y.~L.~S.~Tsai, Q.~Yuan and T.~C.~Yuan,
  JCAP {\bf 1406}, 030 (2014), and reference therein.


\bibitem{Cheung:2013dua}
  C.~Cheung and D.~Sanford,
  JCAP {\bf 1402}, 011 (2014); L. Calibbi, A. Mariotti, and P. Tziveloglou,
  JHEP 10 (2015) 116. 

\bibitem{Cai:2017wdu}
  C.~Cai, Z.~H.~Yu and H.~H.~Zhang,
  Nucl.\ Phys.\ B {\bf 924}, 128 (2017).

\bibitem{Xiang:2017yfs}
  Q.~F.~Xiang, X.~J.~Bi, P.~F.~Yin and Z.~H.~Yu,
  arXiv:1707.03094 [hep-ph].


\bibitem{seesaw}
P. Minkowski, Phys. Lett. B 67, 421 (1977); T. Yanagida, in Proceedings of the Workshop on Unified
Theory and Baryon Number of the Universe, edited by.O.
Sawada and A. Sugamoto (KEK, Tsukuba, Ibaraki 305-
0801 Japan, 1979) p. 95; M. Gell-Mann, P. Ramond, and R. Slansky, in Super-
gravity, P. van Nieuwenhuizen and D. Freedman (eds.),
(Amsterdam: North Holland, 1980).


\bibitem{Cai:2015kpa}
  C.~Cai, Z.~M.~Huang, Z.~Kang, Z.~H.~Yu and H.~H.~Zhang,
  Phys.\ Rev.\ D {\bf 92}, no. 11, 115004 (2015);  C.~Cai, Z.~Kang, Z.~Luo, Z.~H.~Yu and H.~H.~Zhang,
  arXiv:1711.07396 [hep-ph].

\bibitem{Heeck:2015qra}
  J.~Heeck and S.~Patra,
  Phys.\ Rev.\ Lett.\  {\bf 115}, no. 12, 121804 (2015).



\bibitem{Berlin:2016eem}
  A.~Berlin, P.~J.~Fox, D.~Hooper and G.~Mohlabeng,
  JCAP {\bf 1606}, no. 06, 016 (2016).


 \bibitem{Garcia-Cely:2015quu}
  C.~Garcia-Cely and J.~Heeck,
  JCAP {\bf 1603}, 021 (2016).



\bibitem{Cao:2017rjr}
  J.~Cao, X.~Guo, L.~Shang, F.~Wang and P.~Wu,
  arXiv:1712.05351 [hep-ph].


\bibitem{DelNobile:2015bqo}
  E.~Del Nobile, M.~Nardecchia and P.~Panci,
  JCAP {\bf 1604}, no. 04, 048 (2016). 

\bibitem{PM}
S. Dimopolous and H. Georgi, Nucl. Phys. B193 (1981) 150; S. Weinberg, Phys.
Rev. D26 (1982) 287; N. Sakai and T. Yanagida, Nucl. Phys. B197 (1982) 533.

\bibitem{Ibanez:1991hv}
  L.~E.~Ibanez and G.~G.~Ross,
  Phys.\ Lett.\ B {\bf 260}, 291 (1991).

\bibitem{Krauss:1988zc}
  L.~M.~Krauss and F.~Wilczek,
  Phys.\ Rev.\ Lett.\  {\bf 62}, 1221 (1989).
  doi:10.1103/PhysRevLett.62.1221


 \bibitem{Martin:1992mq}
  S.~P.~Martin,
  Phys.\ Rev.\ D {\bf 46}, R2769 (1992)

 \bibitem{Martin:1996kn}
  S.~P.~Martin,
  Phys.\ Rev.\ D {\bf 54}, 2340 (1996).


\bibitem{Kadastik:2009dj}
  M.~Kadastik, K.~Kannike and M.~Raidal,
  Phys.\ Rev.\ D {\bf 81}, 015002 (2010).

\bibitem{Mambrini:2015vna}
  Y.~Mambrini, N.~Nagata, K.~A.~Olive, J.~Quevillon and J.~Zheng,
  Phys.\ Rev.\ D {\bf 91}, no. 9, 095010 (2015).

\bibitem{Frigerio:2009wf}
  M.~Frigerio and T.~Hambye,
  Phys.\ Rev.\ D {\bf 81}, 075002 (2010).


\bibitem{Guo:2015lxa}
  J.~Guo, Z.~Kang, P.~Ko and Y.~Orikasa,
  Phys.\ Rev.\ D {\bf 91}, no. 11, 115017 (2015).

\bibitem{Patra:2016ofq}
  S.~Patra, W.~Rodejohann and C.~E.~Yaguna,
  JHEP {\bf 1609}, 076 (2016); 
  S.~Singirala, R.~Mohanta and S.~Patra,
  arXiv:1704.01107 [hep-ph]; 
  T.~Nomura and H.~Okada,
  arXiv:1705.08309 [hep-ph];   D.~Nanda and D.~Borah,
  Phys.\ Rev.\ D {\bf 96}, no. 11, 115014 (2017).




\bibitem{TuckerSmith:2001hy}
  D.~Tucker-Smith and N.~Weiner,
  Phys.\ Rev.\ D {\bf 64}, 043502 (2001).



  \bibitem{Cui:2017nnn}
  X.~Cui {\it et al.} [PandaX-II Collaboration],
  Phys.\ Rev.\ Lett.\  {\bf 119}, no. 18, 181302 (2017).

\bibitem{Kang:2013wm}
  Z.~Kang, Y.~Liu and G.~Z.~Ning,
  JHEP {\bf 1309}, 091 (2013).



\bibitem{bilepton}
B. O¡¯Leary, W. Porod and F. Staub, JHEP 1205 (2012) 042;  
  L.~Delle Rose, S.~Khalil, S.~J.~D.~King, C.~Marzo, S.~Moretti and C.~S.~Un,
  arXiv:1706.01301 [hep-ph].


\bibitem{iIDM}
  C.~Arina and N.~Fornengo,
  JHEP {\bf 0711}, 029 (2007).

\bibitem{Guo:2013sna}
  J.~Guo, Z.~Kang, T.~Li and Y.~Liu,
  JHEP {\bf 1402}, 080 (2014).


\bibitem{Cao:2017cjf}
  J.~Cao, X.~Guo, Y.~He, L.~Shang and Y.~Yue,
  JHEP {\bf 1710}, 044 (2017).



\bibitem{McKay:2017xlc}
  J.~McKay and P.~Scott,
  arXiv:1712.00968 [hep-ph].


\bibitem{Blinov:2015qva}
  N.~Blinov, J.~Kozaczuk, D.~E.~Morrissey and A.~de la Puente,
  Phys.\ Rev.\ D {\bf 93}, no. 3, 035020 (2016).

\bibitem{Cirelli:2015bda}
  M.~Cirelli, T.~Hambye, P.~Panci, F.~Sala and M.~Taoso,
  JCAP {\bf 1510}, no. 10, 026 (2015).


\bibitem{Chun:2015mka}
  E.~J.~Chun and J.~C.~Park,
  Phys.\ Lett.\ B {\bf 750}, 372 (2015).


\bibitem{Mitridate:2017izz}
  A.~Mitridate, M.~Redi, J.~Smirnov and A.~Strumia,
  JCAP {\bf 1705}, no. 05, 006 (2017).



\bibitem{FERMILAT}
Fermi-LAT, MAGIC Collaboration, M. L. Ahnen et al.,
JCAP 1602 no. 02, (2016) 039. 

\bibitem{HESS}
A. Abramowski et al. [HESS Collaboration], Phys. Rev. Lett. 110, 041301 (2013).

 \bibitem{Bramante:2016rdh}
  J.~Bramante, P.~J.~Fox, G.~D.~Kribs and A.~Martin,
  Phys.\ Rev.\ D {\bf 94}, no. 11, 115026 (2016).

  \bibitem{Chen:2017cqc}
  X.~Chen {\it et al.} [PandaX-II Collaboration],
  arXiv:1708.05825 [hep-ex].


\bibitem{Rodejohann:2015lca}
  W.~Rodejohann and C.~E.~Yaguna,
  JCAP {\bf 1512}, no. 12, 032 (2015); 
  A.~Biswas, S.~Choubey and S.~Khan,
  JHEP {\bf 1608}, 114 (2016); 
  K.~Kaneta, Z.~Kang and H.~S.~Lee,
    JHEP {\bf 1702}, 031 (2017);
  M.~Klasen, F.~Lyonnet and F.~S.~Queiroz,
  Eur.\ Phys.\ J.\ C {\bf 77}, no. 5, 348 (2017); 
  P.~Bandyopadhyay, E.~J.~Chun and R.~Mandal,
  arXiv:1707.00874 [hep-ph].







 \end{thebibliography}
\end{document}